\documentclass[12pt]{article}

\usepackage{paralist}
\usepackage{amsfonts}

\usepackage{amsmath}
\usepackage{amssymb}
\usepackage{amsfonts}

\begin{document}

\newtheorem{theorem}{Theorem}
\newtheorem{corollary}{Corollary}

\fontsize{12}{6mm}\selectfont
\setlength{\baselineskip}{2em}

$~$\\[.35in]
\newcommand{\dss}{\displaystyle}
\newcommand{\raro}{\rightarrow}
\newcommand{\be}{\begin{equation}}

\def\sech{\mbox{\rm sech}}
\def\sn{\mbox{\rm sn}}
\def\dn{\mbox{\rm dn}}
\thispagestyle{empty}

\begin{center}
{\Large\bf Geometric Approaches for Generating }  \\    [2mm]
{\Large\bf Prolongations for Nonlinear Partial Differential Equations }  \\    [2mm]
\end{center}

\vspace{1cm}
\begin{center}
{\bf Paul Bracken}                        \\
{\bf Department of Mathematics,} \\
{\bf University of Texas,} \\
{\bf Edinburg, TX  }  \\
{78541-2999}
\end{center}

\vspace{3cm}
\begin{abstract}
The prolongation structure of a two-by-two problem is 
formulated very generally in terms of exterior
differential forms on a standard representation of Pauli matrices.
The differential system is general
without making reference to any specific equation.
An integrability condition is provided which gives by
construction the equation to be investigated and whose components
involve  the structure constants of an $SU(2)$ Lie algebra.
Along side this, a related, different kind of prolongation, a type of
Wahlquist-Estabrook prolongation, over a closed differential
ideal is discussed and some applications are given.
\end{abstract}

\vspace{2mm}
Keywords: integrable, prolongation, connection, differential system,
fibre bundle, conservation law

\vspace{2mm}
MSCs: 35A30, 32A25, 35C05

\newpage
\begin{center}
{\bf I. INTRODUCTION.}
\end{center}

It is a remarkable fact that large classes of partial
differential equations including nonlinear equations
can be studied from a very general
geometric point of view within the confines of the
formalism of differential geometry$^{1-3}$. 
Not only do the integrable equations possess
infinite numbers of conservation law equations as well
as B\"acklund transformations, it can also be said in many
instances that they can be related to surfaces in, for
example, three space and thereby offer a geometric interpretation
for many of these associated properties$^{4}$. Sasaki$^{1}$ made
the observation that the equations which are the necessary
and sufficient condition for the integrability of a linear
problem of the Ablowitz, Kaup, Newell and Segur$^{5}$ (AKNS)
type to describe pseudosherical surfaces. Geometrically integrable
equations were first considered by Chern and Tenenblat$^{6}$
following in that path. They discovered that, briefly,
a differential equation for a real-valued function describes
pseudospherical surfaces if it turns out to be the necessary
and sufficient condition for the existence of certain smooth
functions such that a set of one-forms defined in terms of
them satisfy the structure equations of a surface of constant
Gaussian curvature.

This is not the only way to regard integrability. There exists
a connection between pseudo-spherical surfaces and integrability
of differential equations that goes well beyond the AKNS 
framework. A differential equation for a real-valued function
is kinematically integrable if it is the integrability condition
of a one-parameter family of linear problems$^{7}$. Another approach
to integrability is the formal symmetry approach studied by
Mikhailov, Shabat and Sokolov$^{8}$. An equation is formally
integrable if it possesses a formal symmetry of infinite rank.

Wahlquist and Estabrook$^{9-10}$ made a significant advance when
they found that prolongations can be constructed for nonlinear
equations. They were interested at first in studying the 
Korteweg-de Vries equation and looking at the associated algebra.
A kind of prolongation over a fibre bundle was found which
corresponds to the Pfaffian system which gives the equation
upon projecting to the transversal integral manifold.
This approach yields results which can be exploited to
develop Lax pairs and to study the B\"acklund properties$^{11}$
of the system, as will be seen here.

The objective here is to find prolongation structures that
can be obtained for a large class of equations given by a
two-by-two problem based on an $SU (2)$ Lie algebra and 
expressed in terms of differential forms. This results
in a geometric approach which does not assume the form
of any specific equation at the outset. The integrability
condition for the Pfaffian system can be expressed as the 
vanishing of a traceless two-by-two matrix of two forms.
This gives by construction the nonlinear equation to be
studied. A prolongation structure for a nonlinear equation
consists of a system of Pfaffian equations for a set of
pseudopotentials, that is functions, which serve as
potentials for conservation laws in a generalized sense.
It will be shown how these prolongations for the two-by-two system
can be derived recursively at first. In the first type of 
prolongation discussed here, forms are used which satisfy an
integrability condition and define a type of connection 
in terms of pseudopotentials.

A prolongation method of a different but related kind is
developed next which is an extension of the Wahlquist
and Estabrook approach$^{9-10,12-14}$. The procedure is
somewhat different from their original approach. 
It relies on finding an underlying Pfaffian system which
constitutes a closed differential ideal and reduces to the
equation on the transversal integral manifold. This can
be regarded as a generalization of the Frobenius Theorem
to establish complete integrability$^{15}$. The method of
prolongation introduces over the base manifold a type
of fibre bundle which is endowed with a Cartan-Ehresmann
connection. The vanishing of the connection form is the
necessary and sufficient condition for the existence of this
type of prolongation. The theorem developed here for this 
operation is very suitable for applications and some of
these will be mentioned further on$^{16-18}$. Thus, in this
second approach, a differential system which gives the
equation on the transversal integral manifold is found,
and this differential system is used to solve for the
quantities which appear in the connection forms. These two
approaches are quite complementary to each other and it
might be of interest to put them together here.

\begin{center}
{\bf II. PROLONGATION STRUCTURE FOR A TWO-BY-TWO PROBLEM.}
\end{center}

Consider the Pfaffian system which is given by
\be
\xi_i =0,
\qquad
\xi_i = d y_i - \Omega_{ij} y_j,
\quad
i,j =1,2.
\label{eq2.1}
\end{equation}
In \eqref{eq2.1}, $\Omega$ is a traceless two-by-two 
matrix which consists of a set of one forms. They can be 
thought of as quite general, but may be taken to constitute
a one-parameter family of forms which, projected onto
the solution manifold, depend on the independent variables, 
the dependent variables and their derivatives. The form
of the matrix of one-forms $\Omega$ is given explicitly as
\be
\Omega = (\Omega_{ij}) = \omega_l \sigma_l =
\begin{pmatrix}
\omega_3  &  \omega_1 - i \omega_2  \\
\omega_1 + i \omega_2  & - \omega_3  \\
\end{pmatrix},
\label{eq2.2}
\end{equation}
where $\sigma_l$, $l=1,2,3$ are the Pauli matrices.
Using \eqref{eq2.2} for $\Omega$, the exterior
differential system in \eqref{eq2.1} takes the form,
\be
\xi_1 = dy_1 -y_1  \omega_3  - y_2 ( \omega_1 -i \omega_2 ) ,
\qquad
\xi_2 = d y_2 - y_1 \, ( \omega_1 +i \omega_2 )  + y_2 \,\omega_3 .
\label{eq2.3}
\end{equation}
The integrability conditions for \eqref{eq2.1} are expressed
as the vanishing of a traceless two-by-two matrix of two-forms
$\Theta$,
\be
\Theta =0,
\qquad
\Theta = d \Omega - \Omega \wedge \Omega.
\label{eq2.4}
\end{equation}
This gives by construction the nonlinear equation which is
of interest. The components of $\Theta$ can be expressed in 
the form,
\be
\Theta = (\Theta_{ij}) = \vartheta_l \sigma_l,
\qquad
\vartheta_l = d \omega_l - i \epsilon_{lmn} \,
\omega_m \wedge \omega_n.
\label{eq2.5}
\end{equation}
In \eqref{eq2.5}, the $\epsilon_{lmn}$ represents the
totally antisymmetric constants of an $SU (2)$ Lie algebra,
which is the case considered now. The nonlinear system to
be considered is specified then by
\be
\Theta =0,
\qquad
\vartheta_l =0,
\quad
l=1,2,3.
\label{eq2.6}
\end{equation}
By exterior differentiation of $\Theta$ in \eqref{eq2.4},
it is found that
\be
d \Theta = \Omega \wedge \Theta - \Theta \wedge \Omega.
\label{eq2.7}
\end{equation}
This establishes that the exterior derivatives of the
two-forms $\{ \vartheta_l \}$ are contained in the ring
generated by the set $\{ \vartheta_l \}$.

It is important to realize that \eqref{eq2.1} and integrability
condition \eqref{eq2.4} are both invariant under the
following type of gauge transformation
\be
{\bf y} \raro {\bf y}' = {\bf Q} {\bf y},
\qquad
\Omega \raro \Omega' = {\bf Q} \Omega {\bf Q} ^{-1} + d {\bf Q} {\bf Q}^{-1},
\qquad
\Theta \raro \Theta' = {\bf Q} \Theta {\bf Q}^{-1}.
\label{eq2.8}
\end{equation}
In \eqref{eq2.8}, ${\bf Q}$ is an arbitrary space-time dependent two-by-two
matrix with determinant one. In other words, the gauge 
transformation of $\Omega$ does not change the solution manifold
of the nonlinear equation. The matrix of one-forms $\Omega$
has the interpretation of being a connection on a gauge field,
the two-form is $\Theta$, a curvature or gauge field strength,
and the closure property \eqref{eq2.7}, a Bianchi identity.

{\bf Theorem 2.1.} The exterior derivatives of the forms
$\xi_1$ and $\xi_2$ have the form,
\be
d \xi_1 =- y_2 (\vartheta_1 - i \vartheta_2) 
- y_1 \vartheta_3 + \omega_3 \wedge \xi_1
+ ( \omega_1 -i \omega_2 ) \wedge \xi_2,
\label{eq2.9}
\end{equation}
\be
d \xi_2 = - y_1 (\vartheta_1 +i \vartheta_2) 
+ y_2 \vartheta_3 + (\omega_1 + i \omega_2) \wedge \xi_1
- \omega_3 \wedge \xi_2.
\label{eq2.10}
\end{equation}
Consequently, the derivatives of $\xi_1$ and $\xi_2$
are contained in the ring of forms spanned by $\{ \vartheta_i \}$
and $\{ \xi_i \}$.

{\bf Proof:} Expressions for the $d y_i$ follow from \eqref{eq2.3},
and from \eqref{eq2.5} $d \omega_i$ can be obtained by writing
$$
d \omega_1 = \vartheta_1 + 2 i \omega_2 \wedge \omega_3,
\qquad
d \omega_2 = \vartheta_2 - 2 i \omega_1 \wedge \omega_3,
\qquad
d \omega_3 = \vartheta_3 + 2 i \omega_1 \wedge \omega_2.
$$
The exterior derivative of $\xi_1$ from \eqref{eq2.3} is
given by
$$
d \xi_1 =-y_1 \vartheta_3 - 2 i y_1 \omega_1 \wedge \omega_2
+ \omega_3 \wedge ( \xi_1 + y_1 \omega_3  + y_2 ( \omega_1 -i \omega_2))
$$
$$
- y_2 ( \vartheta_1 + 2 i \omega_2 \wedge \omega_3 -i \vartheta_2 
- 2 \omega_1 \wedge \omega_3)  + ( \omega_1 -i \omega_2 )
\wedge ( \xi_2 + y_1 ( \omega_1 +i \omega_2 )  - y_2 \omega_3)
$$
$$
= - y_2 \vartheta_1 + i y_2 \vartheta_2 -y_1 \vartheta_3 + \omega_3
\wedge \xi_1 + ( \omega_1 -i \omega_2 ) \wedge \xi_2
$$
$$
- 2 i y_1 \omega_1 \wedge \omega_2 + y_2 \omega_3 \wedge
(\omega_1 -i \omega_2) - 2 i y_2 \omega_2 \wedge \omega_3 + 2 y_2
\omega_1 \wedge \omega_3 + 2 i y_1 \omega_1 \wedge \omega_2
-y_2 ( \omega_1 -i \omega_2 ) \wedge \omega_3.
$$
The second line in the final result vanishes and we are left with
\eqref{eq2.9}. The proof of \eqref{eq2.10} proceeds in the
same way.

{\bf Corollary 2.1.} The exterior derivatives \eqref{eq2.9} and
\eqref{eq2.10} can be expressed in terms of the matrix
elements of $\Omega$ and $\Theta$ in \eqref{eq2.2} and \eqref{eq2.5}
for $i=1,2$ as follows,
\be
d \xi_i =- \Theta_{ij} y_j - \Omega_{ij} \wedge \xi_j.
\label{eq2.12}
\end{equation}

The one-forms $\xi_1$ and $\xi_2$ can be used to generate
an ideal which assumes a standard Riccati form by taking
particular linear combinations of them. The new one-forms
which result are called $\xi_3$ and $\xi_4$, and are
defined by calculating in the following way; first,
$$
y_1^2 \xi_3 = y_1 \xi_2 - y_2 \xi_1 = y_1 d y_2 -y_2 d y_1
- y_1^2 ( \omega_1 +i \omega_2 ) + 2 y_1 y_2 \omega_3
+ y_2^2 ( \omega_1 -i \omega_2 ).
$$
Therefore, $\xi_3$ is given by
\be
\xi_3 = d ( \frac{y_2}{y_1}) - ( \omega_1 + i \omega_2)
+ 2 (\frac{y_2}{y_1}) \omega_3 + ( \frac{y_2}{y_1})^2
( \omega_1 -i \omega_2).
\label{eq2.13}
\end{equation}
In a similar fashion,
$$
y_2^2 \xi_4 = y_2 \xi_1 - y_1 \xi_2 = y_2 d y_1 -y_1 d y_2
- y_2^2 ( \omega_1 -i \omega_2 ) -2 y_1 y_2 \omega_3 +
y_1^2 ( \omega_1 +i \omega_2),
$$
and so $\xi_4$ is given by
\be
\xi_4 = d (\frac{y_1}{y_2}) - ( \omega_1 -i \omega_2)
- 2 (\frac{y_1}{y_2}) \omega_3 + ( \frac{y_1}{y_2})^2
( \omega_1 +i \omega_2).
\label{eq2.14}
\end{equation}
Introducing the new projective variables $\xi_3$ and
$\xi_4$ defined to be
\be
y_3 = \frac{y_2}{y_1},
\qquad
y_4 = \frac{y_1}{y_2}
\label{eq2.15}
\end{equation}
into the expressions for $\xi_3$ and $\xi_4$, Pfaffian
system \eqref{eq2.1} takes the Riccati form,

{\bf Theorem 2.2.} The exterior derivatives of the forms
$\xi_3$ and $\xi_4$ in \eqref{eq2.14}-\eqref{eq2.15} are
given by
\be
d \xi_3 =- ( \vartheta_1 + i \vartheta_2 ) + y_3^2
(\vartheta_1 - i \vartheta_2) + 2 y_3 \vartheta_3
- 2 ( \omega_3 + y_3 ( \omega_1 - i \omega_2 )) \wedge 
\xi_3,
\label{eq2.16}
\end{equation}
\be
d \xi_4 =- ( \vartheta_1 -i \vartheta_2) +y_4^2 ( \vartheta_1
+ i \vartheta_2 ) - 2 y_4 \vartheta_4 + 2 ( \omega_3
- y_4 ( \omega_1 + i \omega_2) ) \wedge \xi_4.
\label{eq2.17}
\end{equation}
Thus, the derivatives of $\xi_3$ and $\xi_4$ given in
\eqref{eq2.14}-\eqref{eq2.15} are contained in the ring of 
forms spanned by $\{ \vartheta_i \}$ and $\{ \xi_i \}$.

Theorem 2.2 is proved along the same lines as the previous
theorem where $d y_3$ and $d y_4$ are obtained from
\eqref{eq2.14} and \eqref{eq2.15} respectively.

The results of Theorem 2.2 can be cast in the general form,
\be
d \xi_3 = \beta_1 \wedge \xi_3 + \sum_{j=1}^3 \, c_{3j} \vartheta_j,
\quad
d \xi_4 = \beta_2 \wedge \xi_4 + \sum_{j=1}^3 \, c_{4 j} \vartheta_j,
\label{eq2.18}
\end{equation}
where the coefficients $c_{ij}$ are function valued quantites 
and the $\beta_i$ are one-forms defined to be
\be
\beta_1 =- 2 \omega_3 -2 y_3 ( \omega_1 -i \omega_2 ),
\qquad
\beta_2 =2 \omega_3 -2 y_4 ( \omega_1 +i \omega_2).
\label{eq2.19}
\end{equation}
It is interesting to note that the exterior derivatives of
the forms $\beta_i$ in \eqref{eq2.19} have the same generic
form as that expressed on the right side of \eqref{eq2.18}.
Based on these results, there are a series of prolongation
results which can be stated and proved along lines similar
to the ones given. These results will be collected together
in Theorem 2.3.

{\bf Theorem 2.3.} $(i)$ Define one-forms $\xi_5$ and $\xi_6$
to have the form,
\be
\xi_5 = d y_5 + 2 \omega_3  + 2 y_3 (\omega_1 -i \omega_2),
\qquad
\xi_6 = d y_6 -2 \omega_3 + 2 y_4 ( \omega_1 +i \omega_2).
\label{eq2.20}
\end{equation}
The exterior derivatives of $\xi_5$ and $\xi_6$ can be
expressed in the form,
\be
d \xi_5 =2 y_3 (\vartheta_1 -i  \vartheta_2) + 2 \vartheta_3
+ 2 \xi_3 \wedge ( \omega_1 -i \omega_2),
\qquad
d \xi_6 = 2 y_4 ( \vartheta_1 +i \vartheta_2) -2 \vartheta_3
+ 2 \xi_4 \wedge ( \omega_1 +i \omega_2).
\label{eq2.21}
\end{equation}

$(ii)$ Define the pair of one-forms
\be
\xi_7 = dy_7 - e^{y_5} ( \omega_1 -i \omega_2),
\qquad
\xi_8 = d y_8 - e^{y_6} ( \omega_1 +i \omega_2).
\label{eq2.22}
\end{equation}
The exterior derivatives of $\xi_7$ and $\xi_8$ are given by
\be
d \xi_7 =- e^{y_5} \, ( \vartheta_1 -i \vartheta_2
+ \xi_5 \wedge (\omega_1 -i \omega_2)),
\qquad
d \xi_8 =- e^{y_6} \, ( \vartheta_1 +i \vartheta_2
+ \xi_6 \wedge ( \omega_1 +i \omega_2)).  \clubsuit
\label{eq2.23}
\end{equation}

Theorem 2.3 is shown along lines identical to that used
for Theorem 2.1 by evaluating exterior derivatives of the
relevant forms, substituting known derivatives and 
then simplifying the resulting expression.

These results are quite significant in that they provide
numerous ways in which to generate conservation laws
explicitly. In fact, an infinite number of them will be 
seen to emerge. 

{\bf A. CONSERVATION LAWS.}

It is worth illustrating how conservation laws can be
generated from the forms which have been produced thus
far. For example, the equations which express $d \beta_1$
and $d \beta_2$ are conservation laws in $1+1$ dimensions.
Under a solution of the original nonlinear equation
$\Theta =0$ with $y_3$, $y_4$ solutions of $\xi_3=0$
and $\xi_4=0$, we can equate to zero $d \beta_i=0$,
$i=1,2$. Expressing the one-form $\beta=\beta_i$ in terms
of the basis set $dx$, $dt$ so that $\beta =
I \, dx +J \, dt$, then $d \beta=0$ implies 
\be
\frac{\partial I}{\partial t} - \frac{\partial J}{\partial x} =0.
\label{eq2.24}
\end{equation}
This signifies that $I$ is a conserved density and
$J$ is a conserved current. In fact $d \beta_i =0$ gives
a one-parameter family of conservation laws as do 
most other results.

By taking a particular structure for the $\omega_i$,
expressions for the conservation laws can be determined
recursively. Suppose that
\be
\omega_1 +i \omega_2 = r \, dx + C\, dt,
\qquad
\omega_1 -i \omega_2 = q \, dx + B \, dt,
\qquad
\omega_3 = \eta \, dx + A\, dt,
\label{eq2.25}
\end{equation}
where $\eta$ is a parameter and $A$, $B$, $C$ are a
one-parameter $\eta$ family of functions of $q$ and $r$.
The one-form $\xi_3$ can be expressed as
$$
\xi_3 = (y_3)_x \, dx + (y_3)_t \, dt
- (r \, dx + C \, dt) +2 y_3 ( \eta \, dx + A \, dt)
+ y_3^2 \, ( q \, dx + B \, dt ).
$$
The coefficients of the basis forms $dx$ and $dt$
yield the equations
$$
(y_3 )_x -r  +2 \eta y_3 + q y_3^2 =0,
\quad
(y_3)_t -C + 2 A y_3 +B y_3^2 =0.
$$
In order to solve this system, substitute $y_3 = \sum_{n=1}^{\infty}
\, \eta^{-n} W_n$ into the first of these equations with 
$W_0=0$ to obtain
$$
-r + W_1 + \sum_{n=1}^{\infty} \, \eta^{-n} ( W_{n,x} 
+ 2 W_{n+1} + q \, \sum_{k=1}^{n-1} \, W_{n-k} W_k )=0.
$$
Equating the coefficients of $\eta^{-n}$ equal to zero,
the following recursion results,
\be
W_1 =r,
\qquad
W_{n,x} + 2 W_{n+1} + q \, \sum_{k=1}^{n-1} \,
W_{n-k} W_k =0.
\label{eq2.26}
\end{equation}
This implies $W_{n+1}$ is a polynomial in $q$, $r$ and
their $x$-derivatives. Moreover, the consistency of
solving $\xi_3=0$ by using only its $x$ dependent part
is guaranteed by complete integrability. Moreover,
the one-form $\beta_1 =-2 ( \omega_3 + ( \omega_1 -i 
\omega_2) y_3 )$ can be expressed in terms of $W_n$ as follows
$$
- \frac{1}{2} \beta_1 = ( \eta + q \sum_{n=1}^{\infty} \,
\eta^{-n} W_n ) \, dx + ( A + B \sum_{n=1}^{\infty} \,
\eta^{-n} W_n ) \, dt.
$$
Consequently, the $n$-th conserved density $I_n$ is given by
$$
I_n = q W_n.
$$
This result would be common to all systems described by
\eqref{eq2.25}. Clearly, the expression of the conserved
current depends on the particular equation considered.
As a remark, notice that by substituting \eqref{eq2.25}
into $\xi_1$ and $\xi_2$, these forms have the structure
$\xi_1 = d y_1 - (\eta y_1 +q y_2) \, dx - (A y_1 + B y_2) \,dt$
and $\xi_2 = d y_2 - ( r y_1 + \eta y_2 ) \, dx 
- ( C y_1 - A y_2 ) \, dt$. This type of structure in a 
prolongation will be exploited in a different setting next.

\begin{center}
{\bf III. A VERSION OF WAHLQUIST-ESTABROOK PROLONGATION.}
\end{center}

A prolongation structure which is different from those
discussed thus far will be developed. This is of the kind used
by Wahlquist and Estabrook$^{9,10}$. Although the pseudopotentials
which will be introduced serve as potentials for conservation
laws in a generalized sense, the kind of results obtained have
other applications and consequences. In the results thus far,
the Pfaffian systems had the property that their exterior
derivatives were contained in a certain ring of forms. 
Similarly, the Pfaffian equations here will reduce to
\be
d \tilde{\xi}_i = \sum_j \, A_{ij} \wedge \tilde{\xi}_j
+ \sum_l \, F_{il} \vartheta_l.
\label{eq3.1}
\end{equation}
The structure of the result in \eqref{eq3.1} can be regarded
as a generalization of the Frobenius condition for complete
integrability of the Pfaffian equation.

Following Wahlquist and Estabrook, consider the manifold
$M = \mathbb R^m$ which has coordinates $(u_1, u_2, \cdots, u_m)$,
with projection map $\pi : M \raro \mathbb R^2$ defined by
$\pi ( u_1, u_2, \cdots, u_m) = ( u_1, u_2 )$. Let there be
defined on $M$ the exterior differential system $\{ \xi_i \}$ 
and denote by ${\cal I} (\xi_i)$ the differential ideal of forms on $M$
generated by $\{ \xi_i \}$. From the point of view of
integral manifolds, the exterior differential system is
completely determined by the associated ideal ${\cal I} (\xi_i)$.
It will be the case that the $\{ \xi_i \}$ are chosen so
that $d {\cal I} \subset {\cal I}$. As a consequence, the Frobenius
Theorem indicates that the Pfaff system $\{ \xi_i \}$
is completely integrable. In the applications which
involve nonlinear equations, the variables $u_1$ and $u_2$
will be identified with the independent variables $x$ and $t$
in the equation, and $u_3$ the dependent variable.
The system $\{ \xi_i \}$ is constructed in such a way that
solutions $u = u (x,t)$ of an evolution equation correspond
to the two-dimensional transversal integral manifolds.

Suppose $N \subset \mathbb R^2$ is coordinatized by the
variables $(x,t)$ and $\pi : M \raro N$, with $s : N \raro M$ 
a cross section of $\pi :M \raro N$. The integral 
manifolds can then be written as sections $S$ in $M$ specified by
\be
s (x,t) = (x,t, u_3 (x,t), \cdots, u_m (x,t)).
\label{eq3.2}
\end{equation}
A bundle can now be constructed based on $M$ so that
$\tilde{M} = M \times \mathbb R^n$. We write
$B = ( \tilde{M}, \tilde{\pi}, M)$ and $\mathbb R^n$ is
coordinatized with coordinates ${\bf y} = ( y_1, \cdots, y_n)$
and whose number at first can be left undetermined. The ${\bf y}$ 
will be referred to as prolongation variables. Everything done so 
far can be lifted up to $\tilde{M}$. Thus, consider the
exterior differential system in $\tilde{M}$ specified by
\be
\tilde{\xi}_i = \tilde{\pi}^{*} \xi_i =0,
\qquad
i=1, \cdots, l,
\qquad
\tilde{\omega}_j =0,
\qquad
j=1, \cdots, n.
\label{eq3.3}
\end{equation}
The forms $\{ \tilde{\omega}_j \}$ have been included in order to
specify a Cartan-Ehresmann connection on $B$.
System \eqref{eq3.3} is called a Cartan prolongation if it
is closed and whenever $S$ is a transversal solution of 
$\{ \xi_i =0 \}$ there should also exist a transverse
local solution $\tilde{S}$ of \eqref{eq3.3} with
$\tilde{\pi} ( \tilde{S}) = S$.
It remains to discuss more carefully the details of the
connection.

The definition 
of connection can be stated in various ways.
A Cartan-Ehresmann connection on $B$ can be regarded as a
field $H$ of horizontal contact elements on $\tilde{M}$
which is supplementary to the field $V$ of the $\tilde{\pi}$-vertical
contact elements. Also $H$ is assumed complete, so every 
complete vector field $X$ on $M$ has a complete horizontal
lift $\tilde{X}$ on $\tilde{M}$. Alternatively, a Cartan-Ehresmann
connection $H$ introduced on $B$ is a system of one-forms
$\{ \tilde{\omega}_i \}$, $i=1, \cdots, n$ with the property
that the mapping $\tilde{\pi}_{*}$ from $H_{\tilde{m}}= \{
\tilde{X} \in T_{\tilde{q}} | \, \tilde{\omega}_i (\tilde{x})=0,
i=1, \cdots, n \}$ onto the tangent space $T_q$ is a bijection
for all $\tilde{q} \in \tilde{M}$. The ideal $\tilde{{\cal I}}$ of
differential forms on $\tilde{M}$, which is generated by
$\tilde{\pi}^{*} {\cal I} \cup H^*$ determines on $\tilde{M}$ the exterior
differential system, which we continue to write as $\{ \xi_i =0\}$.
Here $H^{*}$ is the set of one-forms on $\tilde{M}$ which
vanish on the field $H$.

It remains to specify an explicit expression for the connection
that is considered here. In terms of the coordinates of $B$,
the connection is designated to have the general form
\be
\tilde{\omega}^{k} = d y^{k} - F^k (u_1, \cdots, u_m, {\bf y}) \, dt
- G^{k} ( u_1, \cdots, u_m, {\bf y}) \, dx
\equiv d y^k - \eta^k,
\quad
k=1, \cdots, n.
\label{eq3.4}
\end{equation}
The intention then is to enlarge the differential ideal of
forms by combining the collection $\tilde{\omega}^{k}$ with
the original set of forms, as noted in \eqref{eq3.3}.

The integrability condition requires that the prolonged 
differential ideal $\{ \xi_i, \tilde{\omega}^k \}$ remain
closed. This implies that the differentials of $\tilde{\omega}^k$
can be expressed in the form \eqref{eq3.1},
\be
d \tilde{\omega}^{k} = \sum_{j=1}^{l} \, f^{kj} \xi_j
+ \sum_{j=1}^{n} \, \eta^{k j} \wedge \tilde{\omega}_j.
\label{eq3.5}
\end{equation}
The $f^{kj}$ in \eqref{eq3.5} represent dependent functions
of the bundle coordinates
and $\eta^{ki}$ represent a matrix of one-forms.

For a connection such as \eqref{eq3.4} 
the prolongation condition can be expressed equivalently
using the summation over repeated indices in what follows as,
\be
- d \eta^{i} = \frac{\partial \eta^i}{\partial y^j}
\wedge ( dy^{j} - \eta^{j} ),
\qquad
\mod \tilde{\pi}^{*} ({\cal I}).
\label{eq3.6}
\end{equation}
This  result can be rewritten using the identity
$$
d \eta^{i} = d_M \eta^{i} - ( \frac{\partial \eta^i}
{\partial y^j} ) \wedge d y^{j}.
$$
Here $d_M$ is understood as differentiation with
respect to the variables of the base manifold.
The prolongation condition then becomes,
$$
d_M \eta^i - ( \frac{\partial \eta^{i}}{\partial y^j} )
\wedge \eta^{j} =0,
\qquad
\mod \tilde{\pi}^{*} ({\cal I}).
$$
Introduce the vertical valued one-form as well as the
definitions
$$
\eta = \eta^{i} \frac{\partial}{\partial y^i},
\quad
d \eta = (d_M \eta^{i}) \frac{\partial}{\partial y^{i}},
\quad
[ \eta, \omega ] = ( \eta^{j} \wedge \frac{\partial \omega^i}{\partial y^j}
+ \omega^j \wedge \frac{\partial \eta^i}{\partial y^j} )
\frac{\partial}{\partial y^i}.
$$
The prolongation condition then takes the concise
form
\be
d \eta + \frac{1}{2} [ \eta, \eta ] =0,
\quad
\mod \tilde{\pi}^{*} ({\cal I}).
\label{eq3.7}
\end{equation}
A particular version of connection form \eqref{eq3.4}
of use in generating Lax pairs is, 
\be
\tilde{\Omega}^k = dy^k - \eta^{k} = d y^{k}
- \sum_{i=1}^n \, F^{ki} ( {\bf u}) y^{i} \, dt
- \sum_{i=1}^n G^{ki} ({\bf u}) y^i \, dx.
\label{eq3.8}
\end{equation}
The commutator in \eqref{eq3.7} can be worked out using 
\eqref{eq3.8}. It is given explicitly by
$$
[ \eta, \eta] = ( G^{ji} F^{\nu j} y^{i} \, dx \wedge dt
+ F^{ji} G^{\nu j} y^{i} \, dt \wedge dx
+ F^{ji} G^{\nu j} y^i \, dt \wedge dx
+ G^{ji} F^{\nu j} y^{i} \, dx \wedge dt) \frac{\partial}{\partial y^{\nu}}
$$
\be
= 2 [ F, G ]^{\nu i} y^{i} \frac{\partial}{\partial y^{\nu}} \,
dx \wedge dt.
\label{eq3.9}
\end{equation}
The prolongation condition takes the form
\be
( \frac{\partial F^{\nu i}}{\partial u_j} \, d u_j \wedge dt
+ \frac{\partial G^{\nu i}}{\partial u_j} \, d u_j \wedge dx) y^i
\frac{\partial}{\partial y^{\nu}} + [ F,G]^{\nu i} y^i
\frac{\partial}{\partial y^{\nu}} \, dx \wedge dt =0,
\quad
\mod \tilde{\pi}^{*} ({\cal I}).
\label{eq3.10}
\end{equation}
If the ideal of forms is specified by the system of two forms
$\{ \xi_i \}$ closed over ${\cal I}$, then \eqref{eq3.10} takes the
equivalent form
\be
( \frac{\partial F^{\nu i}}{\partial u_j} \, d u_j \wedge dt
+ \frac{\partial G^{\nu i}}{\partial u_j} \, d u_j \wedge dx )
+ [ F, G ]^{\nu i} \, dx \wedge dt \equiv \lambda^{\nu i}_{j}
\xi^{j}.
\label{eq3.11}
\end{equation}
The objective in any given case is to produce the differential
ideal ${\cal I}$ relevant to the equation and then solve \eqref{eq3.11} 
for the components of the connection $F^{\nu i}$ and $G^{\nu i}$.
In effect, the following theorem has been established$^{7,13}$.

{\bf Theorem 3.1.} Each prolongation of Pfaffian system
$\{ \xi_i =0 \}$ which corresponds to a nonlinear equation on the
integral manifold by a Cartan-Ehresmann connection determines 
a geometrical realization of a Wahlquist-Estabrook partial
Lie algebra ${\cal L}$ by solving \eqref{eq3.11}. Conversely,
every geometrical realization of ${\cal L}$ corresponds to
such a prolongation by constructing \eqref{eq3.4}. Moreover,
on a two-dimensional solution submanifold of the differential 
ideal, the one-forms are annihilated and there exists the 
differential Lax pair
\be
{\bf y}_x =- F {\bf y}, 
\qquad
{\bf y}_t =- G {\bf y}.
\label{eq3.12}
\end{equation}
$$
\clubsuit
$$
The following is an example of a differential system
which gives rise to two important equations.
Let $M = \mathbb R^5$ and let
$\{ u_1, \cdots, u_5 \} = \{ x,t,u,p,q \}$. Define the
following system of two forms to be
\be
\begin{array}{c}
\xi_1 = du \wedge dt -p \, dx \wedge dt,  \\
   \\
\xi_2 = dp \wedge dt - q dx \wedge dt,   \\
   \\
\xi_3 =- du \wedge dx + dq \wedge dx + u \, du \wedge dt
- u \, dq \wedge dt + \beta (u-q) \, du \wedge dt,
\end{array}
\label{eq3.13}
\end{equation}
where $\beta$ in \eqref{eq3.13} is a real, non-zero constant.
Exterior differentiation of the system of $\xi_i$ in
\eqref{eq3.13} yields,
$$
d \xi_1 = dx \wedge \xi_2,
\qquad
d \xi_2 = \frac{1}{u} \, dx \wedge (- \xi_3 +u ((1 + \beta)u
- q ) \xi_1 ),
$$
\be
d \xi_3 = (1- \beta) [ (dq -p \, dx)  \wedge \xi_1 + p \, dt \wedge
\xi_3  ].
\label{eq3.14}
\end{equation}
Clearly, all of the $d \xi_i$ vanish modulo the set of
$\xi_i$ in \eqref{eq3.13}, therefore $d {\cal I} \subset 
{\cal I}$ as required. On the transversal  integral manifold,
it is determined that
\be
\begin{array}{c}
0 = \xi_1|_S = s^{*} \xi_1 = (u_x -p) \, dx \wedge dt,  \\
   \\
0 = \xi_2 |_S = s^* \xi_2 = (p_x -q ) \, dx \wedge dt,  \\
   \\
0 = \xi_3 |_S = s^* \xi_3 = ( u_t -q_t + u (u_x - q_x) 
+ \beta (u-q) u_x ) \, dx \wedge dt.
\end{array}
\label{eq3.15}
\end{equation}
Thus, sectioning of the differential system \eqref{eq3.13}
generates the following system of equations
\be
p= u_x,  \qquad
q = p_x,  \qquad
(u-q)_t + u ( u-q)_x + \beta (u-q) u_x =0.
\label{eq3.16}
\end{equation}
The first two equations in \eqref{eq3.16} imply that
$q= u_{xx}$ and putting this in the third equation,
it is found that
\be
( u - u_{xx} )_t + u (u - u_{xx})_x + \beta (u- u_{xx}) u_x =0.
\label{eq3.17}
\end{equation}
For $\beta=2$, \eqref{eq3.17} becomes the Camassa-Holm equation,
and for $\beta=3$ it takes the form of the Degasperis-Procesi
equation. Solutions of \eqref{eq3.11} for the components of the connection  
based on  systems such as \eqref{eq3.13} have been given$^{16-17}$.

\begin{center}
{\bf IV. SUMMARY.}
\end{center}

It has been seen that a prolongation structure can be obtained
for \eqref{eq2.1} with associated integrability condition \eqref{eq2.4}.
This leads to a geometric formulation of conservation laws, which
can be written down.
Moreover, the two approaches complement each other to a certain degree.
This has also been explored for $SL(2, \mathbb R)$, $O(3)$ and
$SU (3)$, but not reported here. The whole prolongation structure
is gauge covariant, reflecting the gauge invariance of \eqref{eq2.4}.
In fact, although not addressed here, there is a geometrical
application$^{3}$ that can be outlined now. Start
with a general two-dimensional Riemannian
manifold ${\cal M}$. Take an orthonormal basis $\{ e_i \}$ on
the tangent plane $T_p$ at each point $p$. Then the structure
equations for ${\cal M}$ read
\be
dp = \alpha^1 e_1 + \alpha^2 e_2, \quad
d e_1 = \omega e_2,
\quad
d e_2 =- \omega e_1,
\label{eq4.1}
\end{equation}
where $\alpha^1$, $\alpha^2$ are one-forms dual to $\{ e_i \}$
and $\omega$ is the connection one-form. The integrability 
conditions are given by
\be
d \alpha^1 = \omega \wedge \alpha^2,
\quad
d \alpha^2 =- \omega \wedge \alpha^1,
\quad
d \omega =-K \, \alpha^1 \wedge \alpha^2,
\label{eq4.2}
\end{equation}
with $K$ the Gaussian curvature of ${\cal M}$. In this
instance, a surface could arise if we take \eqref{eq2.2}
and identify
\be
\alpha^1 = \omega_2 + \omega_3,
\qquad
\alpha^2 =-2 \omega_1,
\qquad
\omega= \omega_2 - \omega_3.
\label{eq4.3}
\end{equation}
The set of one-forms $(\alpha^1, \alpha^2, \omega)$ satisfying 
these equations describes a surface through these structure equations.
It is remarkable that there is a unification of such
diverse fields under this kind of approach.

\begin{center}
{\bf V. REFERENCES.}
\end{center}

\noindent
$^{1}$ R. Sasaki, Nucl. Phys. {\bf B 154}, 343, (1979).  \\
$^{2}$ R. Sasaki, Proc. R. Soc. Lond., {\bf A 373}, 373, (1980).  \\
$^{3}$ P. Bracken, Int. J. Geom. Meth. Math. Phys., {\bf 6}, 825, (2009).  \\
$^{4}$ N. Kamran and K. Tenenblat, J. Diff. Eqns., {\bf 115}, 75, (1995). \\
$^{5}$ M. J. Ablowitz, D. K. Kaup, A. C. Newell and H. Segur,
Phys. Rev. Letts., {\bf 31}, 125, (1973).   \\
$^{6}$ S. S. Chern and K. Tenenblat, Stud. Appl. Math. {\bf 74}, 55, (1986).  \\
$^{7}$ E. van Groesen and E. M. de Jager, Mathematical structures 
in continuous dynamical systems. Studies in Mathematical Physics, Vol. 6,
North Holland, Amsterdam, (1994). Part II, Ch. 6.  \\
$^{8}$ A. V. Mikhailov, A. B. Shabat and V. V. Sokolov, `The symmetry approach 
to classification of integrable equations', in What is Integrabilty? ed.
V. E. Zakharov, Springer Series in Nonlinear Dynamics (Springer, Berlin,1991),
115-184.   \\
$^{9}$ H. D. Wahlquist and F. B. Estabrook, J. Math. Phys., {\bf 16},
1, (1975).  \\
$^{10}$ H. D. Wahlquist and F. B. Estabrook, J. Math. Phys., {\bf 17},
293, (1976).  \\
$^{11}$ R. Hermann, Phys. Rev. Letts., {\bf 36}, 835, (1976).  \\
$^{12}$ W. F. Shadwick, J. Math. Phys., {\bf 21}, 454, (1980).  \\
$^{13}$ P. Molino, J. Math. Phys., {\bf 25}, 2222, (1980).     \\
$^{14}$ P. Bracken, J. Math. Phys., {\bf 51}, 113502, (2010).  \\
$^{15}$ P. W. Michor, Topics in Differential Geometry, Graduate Studies in Mathematics,
AMS, vol. 93, Providence, RI, (2008).  \\
$^{16}$ P. Bracken, Acta. Appl. Math., {\bf 113}, 247, (2011).  \\
$^{17}$ P. Bracken, Comm. Pure Appl. Math., {\bf 10}, 1345, (2011).  \\
$^{18}$ R. Dodd and A. Fordy, Proc. R. Soc. London, {\bf A 385}, 
389, (1983).  \\
\end{document}